\documentclass[runningheads]{llncs}
\usepackage{graphicx}
\usepackage{amsmath}
\usepackage{amssymb}
\usepackage{textcomp,gensymb}
\usepackage{algorithm}
\usepackage{algpseudocode}
\usepackage{tabularx,booktabs}
\usepackage{pdflscape}
\usepackage{afterpage}
\usepackage{hyperref}
\usepackage{color}
\hypersetup{
  colorlinks   = true, 
  urlcolor     = blue,
  linkcolor    = blue, 
  citecolor    = blue 
}
\usepackage{siunitx}    
\usepackage[T1]{fontenc}
\usepackage{bm}

\usepackage[
    backend=biber,
    style=lncs,
    natbib=true,
    maxnames=3,
    isbn=false,
    url=false,
    eprint=false,
]{biblatex}
\usepackage[super]{nth}
\usepackage{siunitx}
\usepackage[inline]{enumitem}
\usepackage{cleveref}
\crefname{figure}{Fig.}{Figs.}
\Crefname{figure}{Fig.}{Figs.}

\usepackage{dcolumn}
\newcolumntype{d}[1]{D{.}{.}{#1}}
\usepackage{multirow}
\usepackage{upgreek}

\usepackage[dvipsnames,svgnames]{xcolor}
\usepackage[normalem]{ulem}

\usepackage{lipsum}
\usepackage{marvosym}

\usepackage{orcidlink}
\renewcommand{\orcidID}[1]{{\orcidlink{#1}}}

\begin{document}

\title{Ground-truth effects in learning-based fiber orientation distribution estimation in neonatal brains}

\titlerunning{Ground-truth effects on FOD estimation in developing brains}

\author{%
Rizhong Lin\inst{1,2,3,\star}\orcidID{0009-0000-1468-6734} \and
Hamza Kebiri\inst{4,2,}\thanks{R. Lin and H. Kebiri---Equal contribution.}\inst{(\text{\fontsize{8}{10}\Letter})}\orcidID{0000-0001-7592-3166} \and
Ali Gholipour\inst{5,6}\orcidID{0000-0001-7699-4564} \and 
Yufei Chen\inst{3}\orcidID{0000-0002-3645-9046} \and
Jean-Philippe Thiran\inst{1,2,4}\orcidID{0000-0003-2938-9657} \and
Davood Karimi\inst{5}\orcidID{0000-0002-5155-2644}\and
Meritxell Bach Cuadra\inst{4,2}\orcidID{0000-0003-2730-4285}
}

\institute{%
Signal Processing Laboratory (LTS5), \'Ecole Polytechnique Fédérale de Lausanne (EPFL), Lausanne, Switzerland%
\and%
Department of Radiology, Lausanne University Hospital (CHUV) and University of Lausanne (UNIL), Lausanne, Switzerland%
\\\email{hamza.kebiri@unil.ch}%
\and%
College of Electronic and Information Engineering, Tongji University, Shanghai, China%
\and%
CIBM Center for Biomedical Imaging, Switzerland%
\and%
Computational Radiology Laboratory, Department of Radiology, Boston Children’s Hospital and Harvard Medical School, Boston, MA, USA%
\and%
Department of Radiological Sciences, University of California Irvine, CA, USA%
}

\authorrunning{R. Lin and H. Kebiri et al.}

%
\maketitle              
\begin{abstract}
Diffusion Magnetic Resonance Imaging (dMRI) is a non-invasive method for depicting brain microstructure \emph{in vivo}. Fiber orientation distributions (FODs) are mathematical representations extensively used to map white matter fiber configurations. Recently, FOD estimation with deep neural networks has seen growing success, in particular, those of neonates estimated with fewer diffusion measurements. These methods are mostly trained on target FODs reconstructed with \textit{multi-shell multi-tissue constrained spherical deconvolution} (MSMT-CSD), which might not be the ideal ground truth for developing brains. 
Here, we investigate this hypothesis by training a state-of-the-art model based on the U-Net architecture on both MSMT-CSD and \textit{single-shell three-tissue constrained spherical deconvolution} (SS3T-CSD). Our results suggest that SS3T-CSD might be more suited for neonatal brains, given that the ratio between single and multiple fiber-estimated voxels with SS3T-CSD is more realistic compared to MSMT-CSD. Additionally, increasing the number of input gradient directions significantly improves performance with SS3T-CSD over MSMT-CSD. Finally, in an age domain-shift setting, SS3T-CSD maintains robust performance across age groups, indicating its potential for more accurate neonatal brain imaging.


\end{abstract}
\section{Introduction}

Diffusion Magnetic Resonance Imaging (dMRI) is a crucial tool for analyzing \emph{in vivo} the microstructural organization of white matter fibers in the brain. This technique measures the random motion of water molecules, providing unique insights into brain connectivity and revealing abnormalities undetectable by other modalities both \emph{in vivo} and non-invasively. The accurate depiction of white matter (WM) fibers using dMRI is particularly important in early brain development stages, where WM is still in maturation. Advanced methods, such as constrained spherical deconvolution (CSD) \cite{tournier_robust_2007} and multi-shell multi-tissue CSD (MSMT-CSD) \cite{jeurissen2014multi}, are used to reconstruct orientation distribution functions (FODs), enabling both local and global quantitative analyses \cite{descoteaux2008high}.

Supervised deep learning on high-quality datasets have enabled a growing number of methods \cite{lin2019fast,nath2019deep,karimi2021learning,hosseini2022cttrack,kebiri2023deep,jha2023undersampled,spears2023learning,da2024fod} to predict FODs from the raw diffusion signal \cite{karimi2021learning,hosseini2022cttrack,spears2023learning} or its spherical harmonics (SH) representation \cite{lin2019fast,nath2019deep,jha2023undersampled,kebiri2023deep}, or from a spherical deconvolution model \cite{da2024fod}.
Some of these methods \cite{karimi2021learning,kebiri2023robust} have been successfully applied to developing brains for which acquisition times are more constrained, and hence a lower number of measurements is available. In particular, \cite{kebiri2023robust} employed a U-Net architecture to predict large patches of FODs from the newborns of the Developing Human Connectome Project (dHCP), using as few as six diffusion measurements and one b0 as input, making it suitable for clinical acquisitions; \cite{lin_cross-age_2024} investigated age- and site-related domain shifts for fiber estimation within rapidly developing populations. The vast majority of these methods either focus on innovating at the architecture level, the learning choices and parameters, or the input format to the network.

The target ground truth (GT) used to be learned from is not often discussed. The majority of the works \cite{lin2019fast,karimi2021learning,hosseini2022cttrack,kebiri2023deep,jha2023undersampled,spears2023learning,da2024fod} use MSMT-CSD. However, its use for developing brains has been questioned \cite{pietsch2019framework} given the similarity of the response function profiles for each b-value and each tissue at this maturation stage. In fact, the mean signal across b-values of the gray matter (GM) lies within the WM intensities, as opposed to adult data where it lies below, allowing a clear distinction between tissues \cite{pietsch2019framework}. This distinction between tissues is also age-dependent, making the problem of estimating accurate FODs even more challenging for neonates. Recently, \cite{dhollander_improved_2019} has shown that employing the b1000 shell in the case of newborn (dHCP) data with a method relying on a single non-zero b-value measurement and a b0\footnote{$bx$ denotes diffusion weighting, where $x$ is the $b$-value represented in \si{\second\per\square\milli\meter}.}%
, named single-shell three-tissue CSD (SS3T-CSD) \cite{dhollander2016novel,dhollander2016unsupervised}, offers superior detection of crossing fibers compared to MSMT-CSD using all the available shells (b400, b1000 and b2600).

In this work, we explored the suitability of SS3T-CSD for deep learning-based FOD estimation in early development stages. We began by evaluating the consistency of GTs generated by MSMT-CSD, commonly used in this context, and SS3T-CSD, which we investigated here. Next, we extensively tested the state-of-the-art deep learning FOD estimation method on 207 newborn subjects, assessing the impact of training with SS3T-CSD compared to MSMT-CSD GT for the first time. Finally, we examined potential domain shift effects due to age by comparing the performance of models trained and tested on cohorts of similar and different ages using both SS3T-CSD and MSMT-CSD GTs.

\section{Methodology}

\subsection{Datasets and processing}
\label{sec:data} 

We used neonatal brain MRI data from the \nth{3} release of the Developing Human Connectome Project (dHCP) \cite{edwards_developing_2022}, which consists of newborn subjects scanned at the post-menstrual age (PMA) of \numrange{26}{45} weeks using a 3T Philips Achieva scanner. The dMRI data release, having a multi-shell sequence with $b$-values $\qtylist{0;400;1000;2600}{\second\per\milli\meter\squared}$ and \numlist{20;64;88;128} measurements respectively, had been preprocessed using SHARD  \cite{christiaens2021scattered}. The resulting resolution is \qtyproduct{1.5 x 1.5 x 1.5}{\milli\meter}, covering a field of view of \numproduct{100 x 100 x 64} voxels.

Following the approach in \cite{kebiri2023deep}, we generated a WM mask by integrating the \textit{White Matter} and \textit{Brainstem} labels from the provided parcellation of the T2-weighted image of each subject, aligning it to the space of the respective dMRI image using ITK-SNAP \cite{yushkevich2016itk}. Voxels with Fractional Anisotropy (FA) values (computed using MRtrix \cite{tournier2012mrtrix}) greater than 0.25 were also included. 

For our experiments (detailed in \Cref{sec:experiments}), we selected two subsets, denoted as $S_1$ and $S_2$, totaling 207 unique subjects:
\begin{itemize}[topsep=0em]
    \item \textbf{Subset $S_1$}: 100 subjects (PMA at scan: [35.57, 44.29], median: 40.86, mean: 40.11, SD: 2.38).
    \item \textbf{Subset $S_2$} divided into two age groups:
    \begin{itemize}
        \item \emph{Early-stage group} ($S_{2e}$): 65 subjects (PMA at scan: [33.29, 37.86], median: 35.57, mean: 35.69, SD: 1.41).
        \item \emph{Late-stage group} ($S_{2l}$): 65 subjects (PMA at scan: [41.0, 45.14], median: 42.29, mean: 42.40, SD: 1.07).
    \end{itemize}
\end{itemize}

Each subject's data preparation for the deep learning pipeline (described in \Cref{paradigm}) involves sampling single-shell b1000 dMRI images from the full sequence as proposed in \cite{skare2000condition}. These images are normalized by a single b0 image to reduce b-value dependency and then projected onto the corresponding SH spaces to enhance acquisition independence.

\subsection{FOD estimation learning-based model}
\label{paradigm} 
The method \cite{kebiri2023robust} is based on learning a mapping between the SH representation of the raw diffusion signal and the FOD (\Cref{fig:pipeline}, top).

\subsubsection{Ground-truth models.}
MSMT-CSD extends (single-shell single-tissue) CSD \cite{tournier_robust_2007} to accommodate multiple shells and multiple tissues; it aims to solve a constrained linear least squares problem using convex quadratic programming \cite{jeurissen2014multi}, in order to find the optimal FOD coefficients for each tissue, requiring acquisitions with at least 3 b-values.
On the other hand, SS3T-CSD addresses the optimization problem of finding the optimal FOD coefficients differently: 
initially, it fixes prior WM coefficients and estimates GM and cerebrospinal fluid (CSF) coefficients; then, it fixes the CSF coefficients and estimates WM and GM coefficients---this two-step iteration is repeated until convergence \cite{dhollander2016novel},
allowing the method to rely on only two b-value samples (non-b0 and b0). 


\subsubsection{Network specifications.}
The network receives a \numproduct{16x16x16} patch of SH representations from $n_{\text{sig}}$ single-shell diffusion measurements (%
b1000) as input and outputs the corresponding patch of FODs represented in the same SH basis (SH-$L_{\max}$ order 8). We train two models: one targeting FODs estimated from all 300 multi-shell measurements using MSMT-CSD \cite{jeurissen2014multi}, and the other from 88 $b1000$ 
and 20 $b0$ 
measurements using SS3T-CSD \cite{dhollander2016novel}, as depicted upper in \Cref{fig:pipeline}.

\subsubsection{Training.}

For each experimental setting, training was conducted using the Adam optimizer \cite{kingma2014adam} to minimize the $\ell_2$ norm loss of the predicted 45 SH coefficients against the GT FOD SH coefficients, which were generated with MRtrix \cite{tournier2012mrtrix} or one of its forks, MRtrix3Tissue ({\url{https://3Tissue.github.io}}).

During each epoch, 128 patches were randomly extracted from non-empty volumes of each training subject, with each batch containing all 128 patches at a batch size of 1. The training regimen featured an initial learning rate of \num{1e-4}, a dropout rate of 0.1, and used an early stopping strategy. A sliding window technique was employed during inference to process all patches consecutively.

\subsubsection{Implementation and code availability.}

Training was performed on an NVIDIA RTX 2080 Ti GPU, each session lasting approximately 24 hours, using PyTorch \cite{paszke_pytorch_2019}, Lightning \cite{falcon_pytorch_2024}, and MONAI \cite{cardoso_monai_2022}. 
The code is available at \url{https://github.com/Medical-Image-Analysis-Laboratory/dl_fiber_domain_shift.git}.

\subsection{Experiments}\label{sec:experiments}
\begin{figure}[tb]
    \centering
    \includegraphics[width=\textwidth]{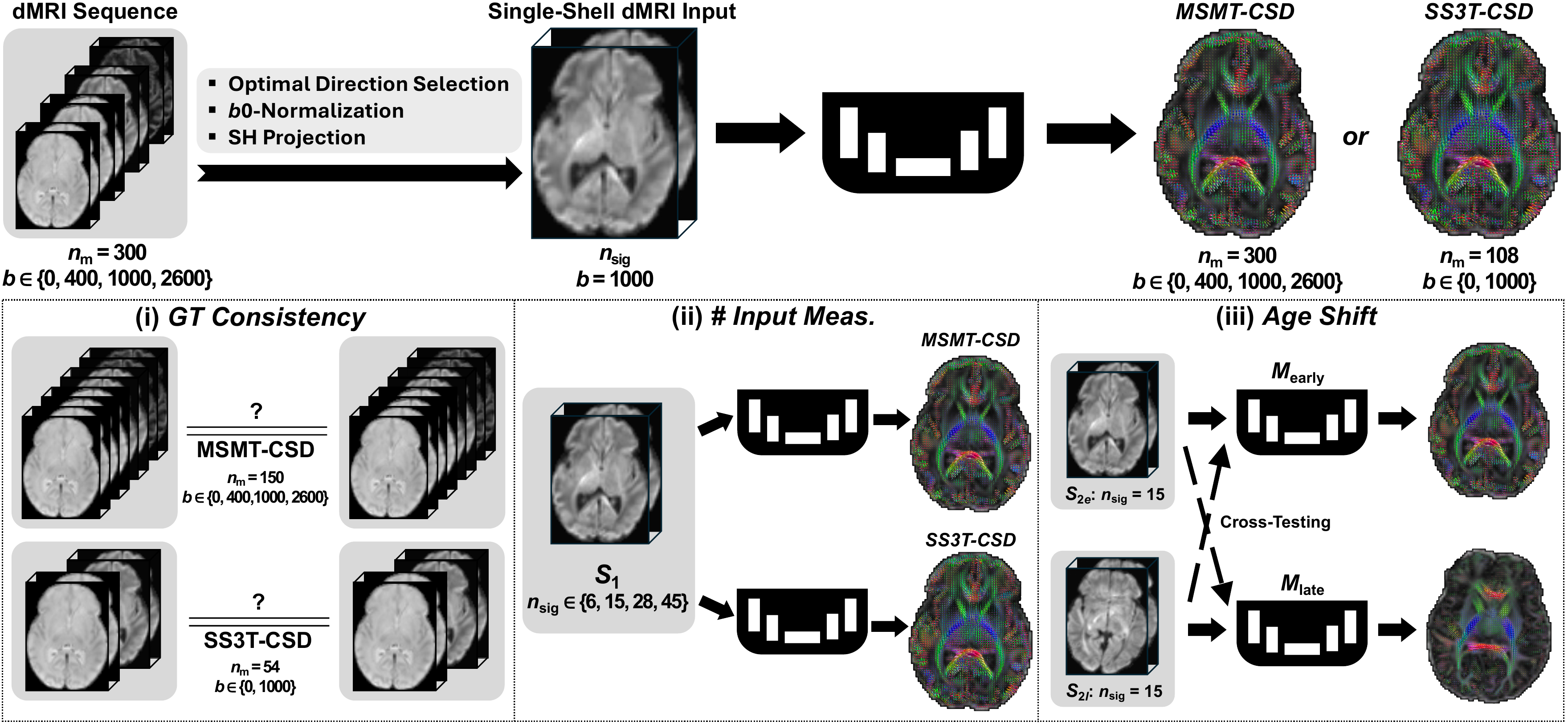}
    \caption{
    Workflow overview. 
    Single-shell b1000 dMRI images are sampled from the full series, normalized by b0 and projected into SH space.
    These images are inputs to a deep learning model predicting GT FODs, using either MSMT-CSD (300 meas.) or SS3T-CSD (108 meas.). The experiments assess: (i) consistency of GT algorithms, (ii) impact of input quantity on model performance, and (iii) model effectiveness across different neonatal developmental stages. Illustrative brain images are from dHCP.
}
    \label{fig:pipeline}
\end{figure}

Experiments testing the efficacy of deep learning models in predicting FODs from MSMT- and SS3T-CSD using dMRI data are outlined in \Cref{fig:pipeline}.

\subsubsection{(i) Ground-truth consistency.}

Following the methodology described in \cite{kebiri2023robust}, we assess GT consistency using MSMT-CSD and SS3T-CSD algorithms on two distinct half subsets of dMRI measurements: 150 measurements each for MSMT-CSD (10 b0's, 32 b400's, 44 b1000's, and 64 b2600's) and 54 measurements each for SS3T-CSD (10 b0's and 44 b1000's). This method evaluates the reliability of the GT algorithms by comparing FODs generated from equivalent but separate subsets%
, using 80 subjects randomly sampled from data subset $S_1$.

\subsubsection{(ii) Model assessment and ablation study on the number of input directions.}
\label{sec:num_of_meas}
Performance between models trained on SS3T-CSD and those trained on MSMT-CSD was compared. This comparison was also performed across a different number of input directions to the network. The network input has $n_{\text{sig}}$ measurements ($n_{\text{sig}} \in \left\{6, 15, 28, 45\right\}$) from the 
b1000
shell, each projected onto an SH basis with $L_{\max}\in\left\{2,4,6,8\right\}$.
In this phase, the data subset $S_1$ was divided into groups of 70 for training, 15 for validation, and 15 for testing.

\subsubsection{(iii) Age-related domain shifts.}

To explore age-related variations within each dataset, we conducted age-specific training and testing within and across these ages \cite{lin_cross-age_2024}. 
As mentioned in \Cref{sec:data}, in the dHCP dataset, subjects in S2 were selected and divided into two age groups, 
denoted as \emph{early} and \emph{late}, respectively. Each age group consists of 65 subjects, further partitioned into splits of 40 for training, 10 for validation, and 15 for testing. We fixed the number of input measurements to 15 (corresponding to SH order 4) and trained separate models on the \emph{early} and \emph{late} age groups. Each model was then tested on its own age group (self-testing) as well as on the other age group (cross-age testing).

\subsection{Evaluation metrics}

Quantitative validation was performed based on the agreement rate (AR) \cite{kebiri2023robust} between the number of fibers estimated by the number of peaks extracted using Dipy \cite{garyfallidis2014dipy} (with a mean separation angle of 45\textdegree, a maximum number of 3 peaks and relative peak threshold of 0.5), the angular error (AE) between those peaks \cite{kebiri2023robust}, and the apparent fiber density (AFD) \cite{raffelt2012apparent} error. For the AFD error, we report the mean absolute percentage error (MAPE) as the magnitude of the SS3T-CSD-derived FODs is higher than that of MSMT-CSD.

\section{Results}
\begin{figure}[tb]
    \centering
    \includegraphics[width=0.8\linewidth]{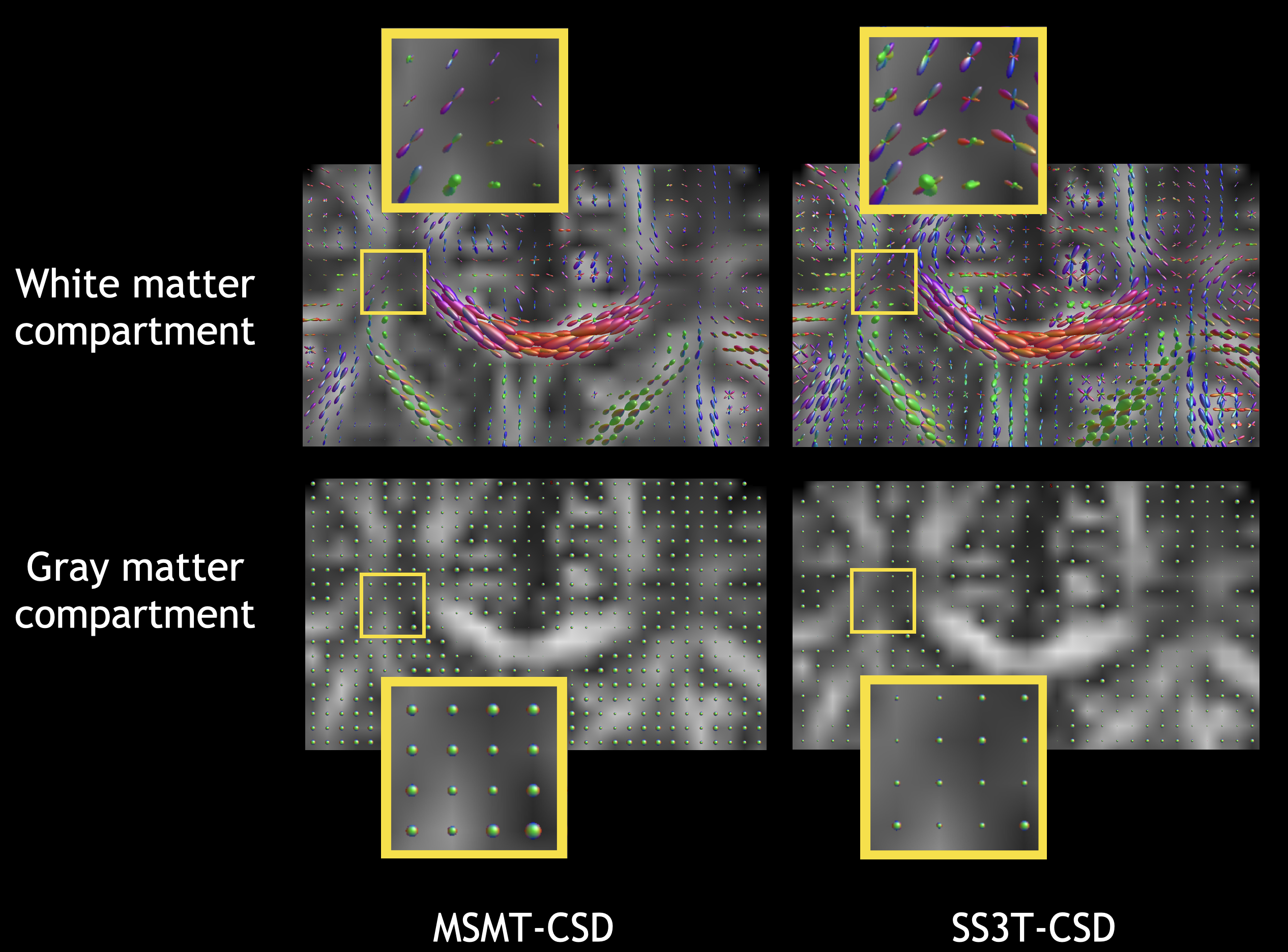}
    \caption{Qualitative comparison of coronal slices of the FODs used as GT for the model, reconstructed with MSMT-CSD and SS3T-CSD, respectively, from the dMRI scan of a subject at 40 weeks PMA. Both gray matter and white matter compartments are displayed on the fractional anisotropy (FA) image computed from dMRI data. FOD estimation and visualization were performed with MRtrix \cite{tournier2012mrtrix}.}
\label{fig:fod}
\end{figure}

Firstly, we show a qualitative comparison between the FOD reconstructed using the two GT methods for a neonate with 40 weeks PMA (\Cref{fig:fod}), where we clearly see missing crossings in MSMT-CSD, likely due to the overestimation of the gray matter compartment, compared to SS3T-CSD.

\subsection{Ground-truth consistency}

\Cref{tab:gs} provides a comparison between the two GT methods. MSMT-CSD shows a higher agreement rate for single fibers but similar angular errors when compared to SS3T-CSD. In contrast, in the 2-fiber configuration, SS3T-CSD outperforms MSMT-CSD in agreement rate and exhibits slightly better angular errors. It is fair to mention that SS3T-CSD employs the minimum number of directions given the SH-order of the FOD (44 directions for 45 coefficients), compared to 140 directions by MSMT-CSD. Hence, the SS3T-CSD consistency errors reported here represent the upper bounds of expected errors. This is particularly relevant for 3-fiber configurations, which are more susceptible to noise and thus less reliable. Importantly, 
closer examination of the confusion matrices (\Cref{tab:confmat}) reveals that
the proportion of multiple fibers estimated by MSMT-CSD is around 23\%, significantly lower than the literature-reported values of over 60\% \cite{jeurissen2013investigating, schilling2022prevalence}, 
whereas SS3T-CSD estimates this proportion to be 61\%.

\begin{table}[tb]
\centering
\caption{Quantitative comparison of the consistency of GT FODs generated with MSMT-CSD and SS3T-CSD. Agreement Rate (AR, \%) and Angular Error (AE, $^\circ$) for each fiber number configuration, and AFD MAPE ($\Updelta$AFD, \%) are listed, together with the number of measurements ($n_{\text{m}}$) used and the corresponding $b$-values. AE is computed among fibers with agreed peak predictions only.}
\label{tab:gs}
\resizebox{0.95\textwidth}{!}{%
\begin{tabular}{|c|c|p{2em}<\centering|cp{2.5em}<\centering|cp{2.5em}<\centering|cp{2.5em}<\centering|c|}
\hline
\multirow{2}{*}{\textbf{Method}} &
  \multirow{2}{*}{\textbf{$\boldsymbol{b}$-values}} &
  \multirow{2}{*}{\textbf{$\boldsymbol{n}_{\text{m}}$}} &
  \multicolumn{2}{c|}{\textbf{1-Fiber}} &
  \multicolumn{2}{c|}{\textbf{2-Fiber}} &
  \multicolumn{2}{c|}{\textbf{3-Fiber}} &
  \multirow{2}{*}{\textbf{$\Updelta$AFD}} \\ \cline{4-9}
 &
   &
   &
  \multicolumn{1}{p{2.5em}<\centering|}{\textbf{AR}} &
  \textbf{AE} &
  \multicolumn{1}{p{2.5em}<\centering|}{\textbf{AR}} &
  \textbf{AE} &
  \multicolumn{1}{p{2.5em}<\centering|}{\textbf{AR}} &
  \textbf{AE} &
   \\ \hline
MSMT-CSD &
  $\left\{0, 400,1000,2600\right\}$ &
  150 &
  \multicolumn{1}{c|}{88.8} &
  6.92 &
  \multicolumn{1}{c|}{45.6} &
  14.91 &
  \multicolumn{1}{c|}{52.8} &
  25.96 &
  2.87 \\ \hline
SS3T-CSD &
  $\left\{0, 1000\right\}$ &
  54 &
  \multicolumn{1}{c|}{63.9} &
  6.76 &
  \multicolumn{1}{c|}{57.7} &
  12.32 &
  \multicolumn{1}{c|}{42.1} &
  23.38 &
  0.43 \\ \hline
\end{tabular}%
}
\end{table}

\begin{table}[tb]
\centering
\caption{Confusion matrices in the number of peaks agreement, normalized over all populations (in \%) for MSMT-CSD and SS3T-CSD ground truth consistency.}
\label{tab:confmat}
\resizebox{0.65\textwidth}{!}{%
\begin{tabular}{clllcclll}
\multicolumn{4}{c}{\textbf{MSMT-CSD}} &
  \qquad{} &
  \multicolumn{4}{c}{\textbf{SS3T-CSD}} \\ \cline{1-4} \cline{6-9} 
\multicolumn{1}{|c|}{\textbf{\# Fibers}} &
  \multicolumn{1}{c|}{\textbf{1}} &
  \multicolumn{1}{c|}{\textbf{2}} &
  \multicolumn{1}{c|}{\textbf{3}} &
  \multicolumn{1}{c|}{} &
  \multicolumn{1}{c|}{\textbf{\# Fibers}} &
  \multicolumn{1}{c|}{\textbf{1}} &
  \multicolumn{1}{c|}{\textbf{2}} &
  \multicolumn{1}{c|}{\textbf{3}} \\ \cline{1-4} \cline{6-9} 
\multicolumn{1}{|c|}{\textbf{1}} &
  \multicolumn{1}{l|}{0.715} &
  \multicolumn{1}{l|}{0.0446} &
  \multicolumn{1}{l|}{0.0052} &
  \multicolumn{1}{c|}{} &
  \multicolumn{1}{c|}{\textbf{1}} &
  \multicolumn{1}{l|}{0.2955} &
  \multicolumn{1}{l|}{0.0775} &
  \multicolumn{1}{l|}{0.0062} \\ \cline{1-4} \cline{6-9} 
\multicolumn{1}{|c|}{\textbf{2}} &
  \multicolumn{1}{l|}{0.0362} &
  \multicolumn{1}{l|}{0.1013} &
  \multicolumn{1}{l|}{0.021} &
  \multicolumn{1}{c|}{} &
  \multicolumn{1}{c|}{\textbf{2}} &
  \multicolumn{1}{l|}{0.0774} &
  \multicolumn{1}{l|}{0.3517} &
  \multicolumn{1}{l|}{0.0494} \\ \cline{1-4} \cline{6-9} 
\multicolumn{1}{|c|}{\textbf{3}} &
  \multicolumn{1}{l|}{0.0035} &
  \multicolumn{1}{l|}{0.0188} &
  \multicolumn{1}{l|}{0.0544} &
  \multicolumn{1}{c|}{} &
  \multicolumn{1}{c|}{\textbf{3}} &
  \multicolumn{1}{l|}{0.0056} &
  \multicolumn{1}{l|}{0.0532} &
  \multicolumn{1}{l|}{0.0835} \\ \cline{1-4} \cline{6-9} 
\end{tabular}%
}
\end{table}

\subsection{Model assessment and number of input directions effect}\label{sec:num_of_directions}
\begin{figure}[htb]
    \centering
    \includegraphics[width=\textwidth]{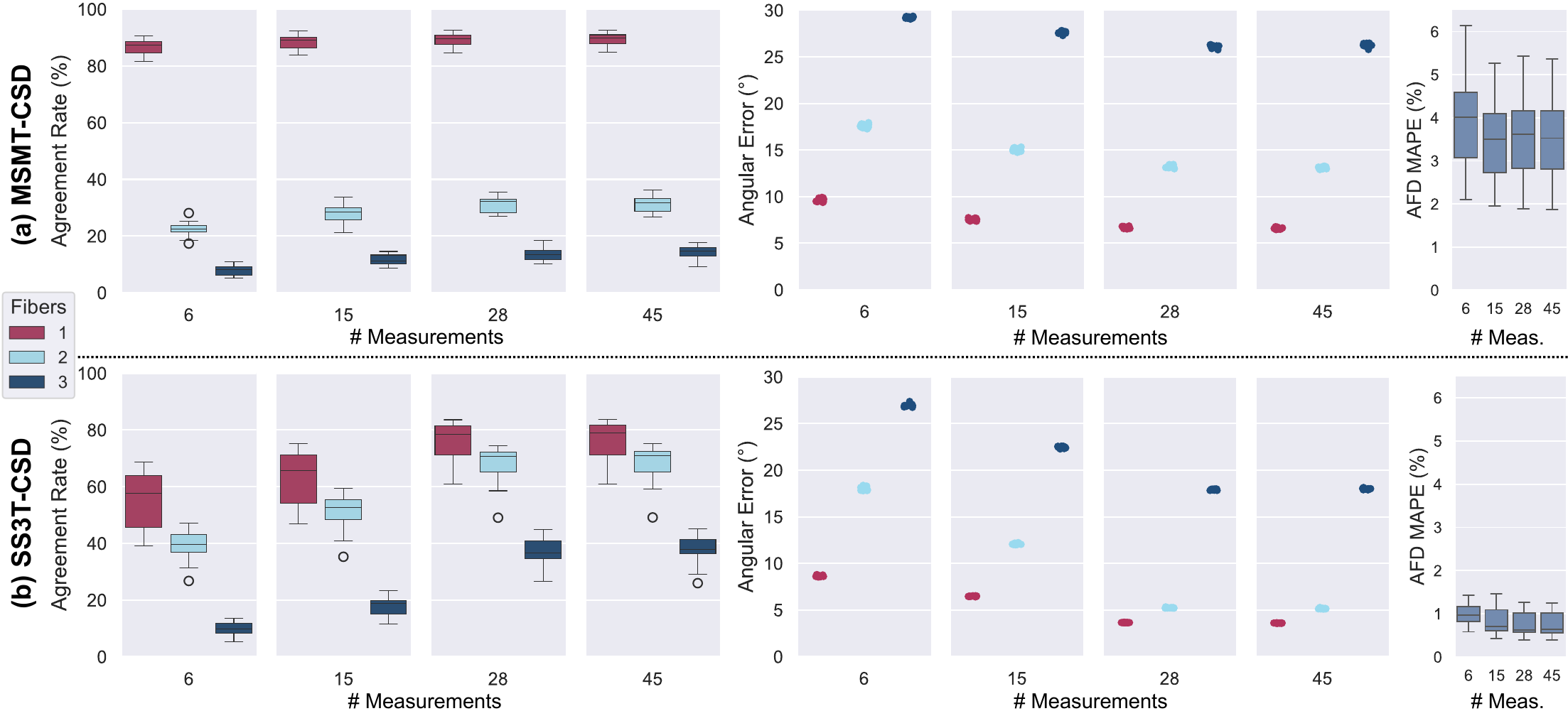}
    \caption{Comparison of performance of the models based on incrementing numbers of input directions, using GT: (a) MSMT-CSD; (b) SS3T-CSD. AR and AE under different fiber number configurations and AFD Error are depicted. 
    }
    \label{fig:dir}
\end{figure}

\Cref{fig:dir} illustrates the comparative results in terms of agreement rate, angular error, and AFD error across varying numbers of input gradient directions, using MSMT-CSD and SS3T-CSD GTs, respectively.

We observe that SS3T-CSD consistently shows higher agreement rates and lower angular errors for multiple fibers compared to MSMT-CSD across all input directions, except for 6 directions where they are comparable. Increasing the number of input directions improves the network's accuracy in predicting multiple fibers more significantly with SS3T-CSD than MSMT-CSD. Specifically, SS3T-CSD achieves an agreement rate of around 70\%--76\% for single and 2-fiber populations with 28--45 input directions, while MSMT-CSD's 2-fiber population agreement rate remains around 30\% compared to 88\% for single fibers, even with 45 input directions.
However, for single fiber populations, MSMT-CSD produces more accurate predictions. For AFD, SS3T-CSD shows lower errors across all input levels compared to MSMT-CSD, with the former exhibiting lower error variance. This can be attributed to the fact that AFD is computed using the same b-value as the input b-value measurement to the network for SS3T-CSD, whereas it is computed with all three shells in MSMT-CSD.

\subsection{Age domain shift}

\begin{figure}[htb!]
    \centering
    \includegraphics[width=\textwidth]{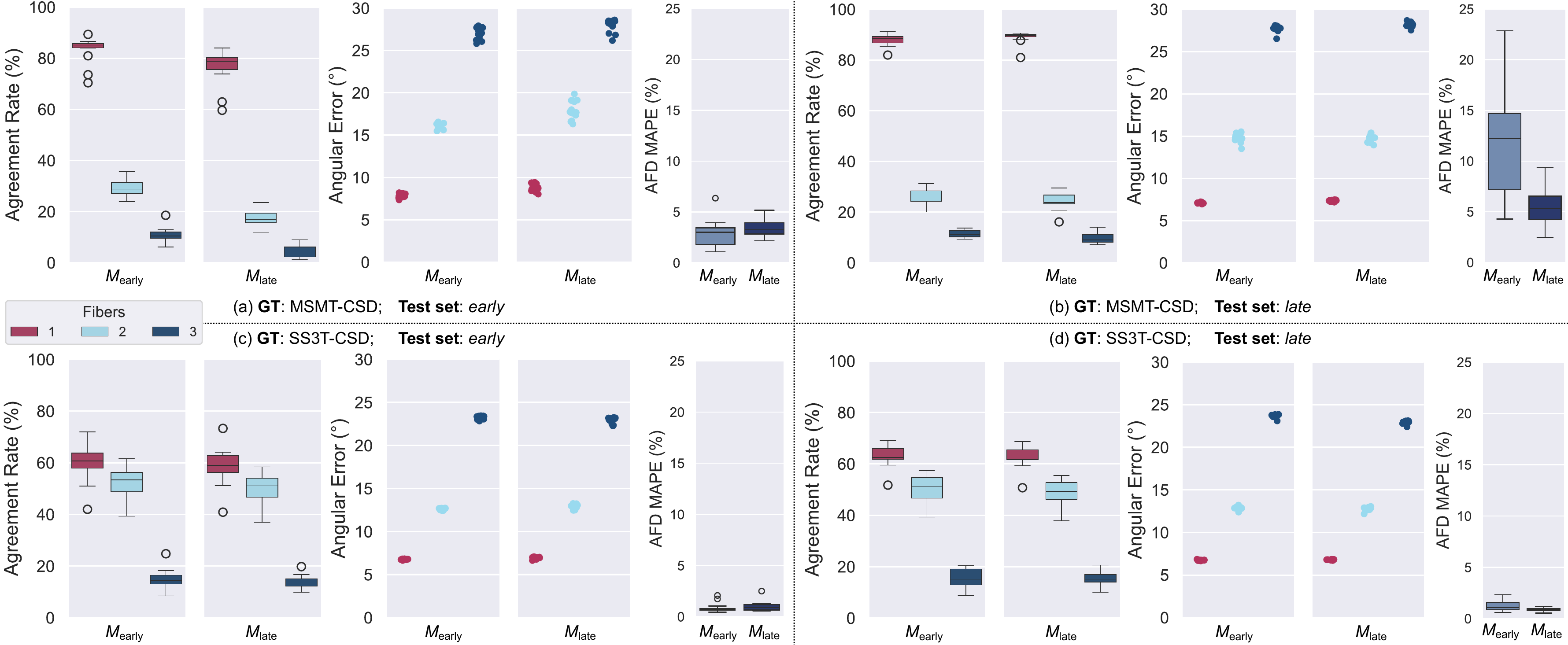}
    \caption{Comparison of performance metrics for FOD estimation models trained and tested on \emph{early} and \emph{late} developmental stages using MSMT-CSD and SS3T-CSD ground truths. The agreement rate, angular error, and AFD error are depicted across models trained on \emph{early} ($M_{\text{early}}$) and \emph{late} ($M_{\text{late}}$) age groups.}
    \label{fig:age}
\end{figure}

\Cref{fig:age} presents the effects of age-specific training and testing on model performance using MSMT- and SS3T-CSD GTs.
We first notice that when testing in the \emph{early}-stage cohort, SS3T-CSD cross-age training is more robust to domain shift across all metrics, including single-fiber populations AR and AE, compared to MSMT-CSD. In fact, the drop for instance in 2-fibers agreement rate goes from around 28\% to 18\% for the latter, but only from 55\% to 50\% for the former. However, for the configuration when tested in the \emph{late}-stage cohort, no significant difference can be observed between the two GTs, except for AFD where SS3T-CSD seems more robust to domain shift. In general, an increased variance is observed in cross-testing scenarios in for both SS3T-CSD- and MSMT-CSD-trained models, especially for AFD.

\section{Conclusion}
We have demonstrated the differences in the training of deep learning models using MSMT-CSD and SS3T-CSD for FOD estimation in neonatal brains. Compared to MSMT-CSD, SS3T-CSD improved accuracy and reliability in multiple fiber configurations, which is a major bottleneck in diffusion MRI and, in particular, for rapidly developing brains \cite{calixto2024white}. However, there is a drop in performance in single-fiber populations compared to MSMT-CSD, potentially due to higher values in gray matter compartments in GT MSMT-CSD that smoothens the overall estimated FODs, resulting in less predicted multi-fibers voxels, and vice-versa for SS3T-CSD and multiple fibers predictions. More b1000 measurements for SS3T-CSD can potentially lower this effect. Furthermore, SS3T-CSD showed robust performance across different age groups, reducing the performance drop caused by age domain shifts. 

Future work will focus on evaluating SS3T-CSD in other pediatric populations, such as fetuses and babies, and exploring downstream performance in tractography. We will also investigate a variety of models beyond U-Net and different strategies to address domain shift challenges through data harmonization and domain adaptation techniques.

\begin{credits}
\subsubsection{Ethics Statement.}

This study uses pre-approved dHCP data by UK Health Research Authority (REC reference: 14/LO/1169), requiring no additional ethical approval.

\subsubsection{\ackname} 
We acknowledge access to the facilities and expertise of the CIBM Center for Biomedical Imaging, a Swiss research center of excellence founded and supported by 
Lausanne University Hospital (CHUV), University of Lausanne (UNIL), \'Ecole Polytechnique Fédérale de Lausanne (EPFL), University of Geneva (UNIGE),  Geneva University Hospitals (HUG),
and the Leenaards and Jeantet Foundations. 

This research was partly supported by grants from the Swiss National Science Foundation (182602 and 215641); 
the US National Institutes of Health (NIH), including: the National Institute of Neurological Disorders and Stroke (R01NS106030 and R01NS128281), the National Institute of Biomedical Imaging and Bioengineering (R01EB032366), and the Eunice Kennedy Shriver National Institute of Child Health and Human Development (R01HD110772); 
and the National Natural Science Foundation of China (62173252).
The opinions expressed herein are solely those of the authors and do not necessarily reflect those of the funding agencies.

\subsubsection{\discintname}
The authors declare no relevant competing interests regarding the content of this article. 

\end{credits}

%
%

\printbibliography

\end{document}